\documentclass[twocolumn,footinbib,floatfix,aps,10pt,pra,reprint]{revtex4-1}
\usepackage{bm}
\usepackage{ulem}
\usepackage{epsfig}
\usepackage{graphicx}
\usepackage{amssymb,amsmath,amsbsy,amsgen,amsfonts}
\usepackage{dcolumn}
\usepackage{amsthm}
\usepackage{mathrsfs}
\usepackage{latexsym}
\usepackage{array}
\usepackage{amstext}
\allowdisplaybreaks[1]
\usepackage{txfonts}

\usepackage{xcolor}

\newcommand{\bra}[1]{\mathrm \langle{#1}|}

\newcommand{\ket}[1]{\mathrm |{#1}\rangle}

\def\bea{\begin{eqnarray}}
\def\eea{\end{eqnarray}}

\begin{document}
\title{Multi-color quantum control for suppressing ground state coherences\\in two-dimensional electronic spectroscopy}
\author{J. Lim$^{1}$}\thanks{These authors contributed equally to this work.}
\author{C.~M. B\"{o}sen$^{1}$}\thanks{These authors contributed equally to this work.}
\author{A.~D. Somoza$^{1}$}
\author{C.~P. Koch$^{2}$}
\author{M.~B. Plenio$^{1}$}\email{martin.plenio@uni-ulm.de}
\author{S.~F. Huelga$^{1}$}\email{susana.huelga@uni-ulm.de}
\affiliation{$^{1}$ Institut f\"{u}r Theoretische Physik and IQST, Albert-Einstein-Allee 11, Universit\"{a}t Ulm, 89081 Ulm, Germany}
\affiliation{$^{2}$ Theoretische Physik, Universit\"{a}t Kassel, Heinrich-Plett-Strasse 40, 34132 Kassel, Germany}

\begin{abstract}

The measured multi-dimensional spectral response of different light harvesting complexes exhibits oscillatory features which suggest an underlying coherent energy transfer. However, making this inference rigorous is challenging due to the difficulty of isolating excited state coherences in highly congested spectra. In this work, we provide a coherent control scheme that suppresses ground state coherences, thus making rephasing spectra dominated by excited state coherences. We provide a benchmark for the scheme using a model dimeric system and numerically exact methods to analyze the spectral response. We argue that combining temporal and spectral control methods can facilitate a second generation of experiments that are tailored to extract desired information and thus significantly advance our understanding of complex open many-body structure and dynamics.

\end{abstract}

\maketitle

%%%%%%%%%%%%%%%
%%%%%%%%%%%%%%% Introduction
%%%%%%%%%%%%%%%

%\section{Introduction}
The concept of excitation energy transfer between donor and acceptor molecules is essential for the elucidation of fundamental transport phenomena in interacting many-body systems~\cite{MayKuehn2003}. Depending on the nature and strength of the system interaction, the effect of the surrounding environment and the considered timescale, the energy transfer can be well described as an incoherent process resulting in hopping kinetics~\cite{forster1946} or it may display coherent features as a result of the formation of delocalized excitons~\cite{Amerongen2000,Valkunas}. Multi-dimensional spectroscopy, which applies sequential short laser pulses with controllable time separation, is particularly well suited for the characterization of energy transfer channels and the observation of coherent features of transport dynamics~\cite{JonasARPC2003,BrixnerJCP2004}. Specifically, by correlating excitation and detection frequencies as a function of the time delay, two-dimensional (2D) electronic spectra are obtained, which can exhibit cross-peaks where the two frequencies differ, indicating electronic coupling between subsystems and associated energy transfer. Varying the time delay, it is possible to monitor energy transfer paths, estimate the corresponding transfer rates and discriminate coherent from incoherent processes. The application of these techniques to the study of photosynthetic membrane pigment-proteins complexes (PPCs)~\cite{Blankenship2002} has shown that the spectral response contains multiple oscillatory features~\cite{EngelN2007,EngelPNAS2010,ColliniN2010,RomeroNP2014,FullerNP2014} whose origin and implications for the description of the system's dynamics are the subject of vigorous discussion (See \cite{Huelga2013,Jonas2018,rmp2018} for recent reviews). The fact that oscillating 2D signals may not only originate from coherent motions in the excited state potential, but could also be induced by vibrational motions in the ground state has made the identification of coherent excited state features a challenging task~\cite{Huelga2013,Jonas2018,rmp2018,PriorPRL2010,ChristenssonJPCB2012,KreisbeckJPCL2012,ChinNP2013,Plenio2013,TiwariPNAS2013,ChenuSR2013,ButkusJCP2014,RomeroNP2014,FullerNP2014,LimNC2015,DeSioNC2016}. The most recent experiments~\cite{Maiuri2018,Donatas2018} using the Fenna-Matthew-Olson (FMO) complex seem to favour a mixed origin of the observed coherences, resulting from coherent electronic-vibrational (vibronic) motions. Previous experiments analyzing charge separation in the PSII reaction centers~\cite{RomeroNP2014,FullerNP2014} were also consistent with an underlying vibronic model.
However, there exist other experiments which advocate a different origin for the observed coherence in FMO, being it purely electronic~\cite{EngelN2007,EngelPNAS2010,EngelChem2018} or even purely vibrational~\cite{Miller2017}. When considering other PPCs, such as harvesting units present in cryptophyte algae, recent work supports non-trivial vibronic dynamics~\cite{Scholes2016}, with experiments showing a correlation between measured transfer rates and vibronic coupling~\cite{Scholes2018}. Discrepancies also extend to the theoretical modelling, where some analyses argue for the compatibility of current observations with a coherent transport of excitations~\cite{Kolli2012}, while others advocate an incoherent transport model~\cite{Aspuru2018}. It is therefore highly desirable to design and perform new experimental tests that can shed further light on the characteristics of the excited state manifold. To suppress unwanted ground state coherences from isotropic samples, a polarization-controlled 2D scheme has been experimentally implemented~\cite{Donatas2012,Donatas2018}, although its performance is degraded when vibronic mixing is present~\cite{refSM}. Broadband pump probe and transient grating schemes, which have also been proposed in the literature~\cite{Joel,Joel2}, do not provide excitation frequency resolution in contrast to the 2D schemes~\cite{refSM}. In this work, we provide a coherent control scheme where a multi-color pulse sequence suppresses the generation of ground state coherences in rephasing spectra, and therefore results in oscillatory 2D signals that are dominated by excited state coherences.

\begin{figure}
	\includegraphics[width=0.45\textwidth]{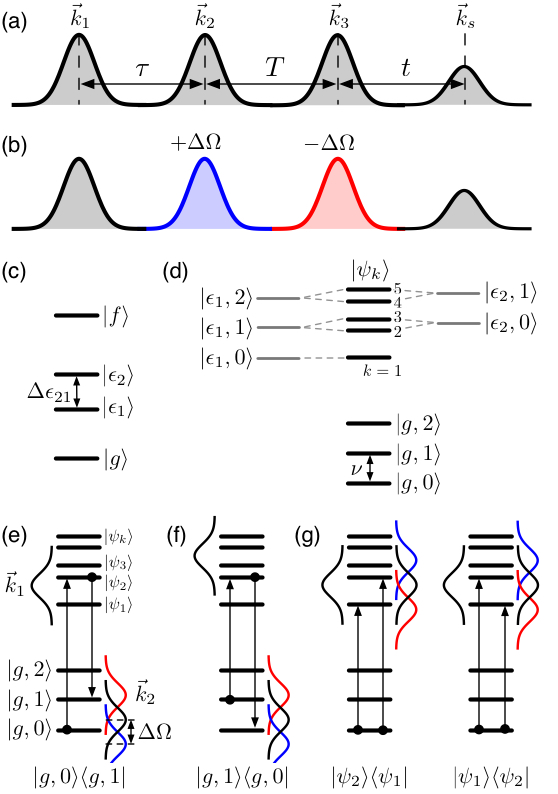}
	\caption{(a) In a single-color scheme, a sample is excited by three pulses and then generates signal, with time intervals denoted by coherence $\tau$, waiting $T$ and rephasing $\tau$ times. (b) A multi-color scheme for suppressing ground state coherences where the second pulse is blue-shifted by $\Delta\Omega>0$ from the first pulse, while the third pulse is red-shifted by $\Delta\Omega$. (c) Electronic eigenstates of a dimer. (d) Energy-level structure of a vibronic model including ground $\ket{g,n}$ and excited $\ket{\psi_k}$ states. The generation of ground state coherences from (e) global ground state $\ket{g,0}$ and (f) vibrationally excited ground state $\ket{g,1}$, and (g) excited state coherences $\ket{\psi_2}\bra{\psi_1}$ and $\ket{\psi_1}\bra{\psi_2}$ between vibronic eigenstates, induced by the first two pulses in 2D measurements.}
	\label{fig:1}
\end{figure}

In 2D electronic spectroscopy, a molecular sample interacts with three laser pulses in sequence, and the resultant third-order molecular polarization generates nonlinear signals~\cite{JonasARPC2003,BrixnerJCP2004}, as schematically shown in Fig.~\ref{fig:1}(a). In non-collinear 2D measurements, where the laser pulses propagate along different directions, described by wave vectors $\vec{k}_1$, $\vec{k}_2$, $\vec{k}_3$, the signals are emitted along several phase-matched directions $\vec{k}_{\rm s}=\pm \vec{k}_1 \pm \vec{k}_2 \pm \vec{k}_3$. Rephasing spectra, measured at $\vec{k}_{\rm s}=-\vec{k}_1 +\vec{k}_2 +\vec{k}_3$, are obtained by Fourier transforming the optical response with respect to the time intervals between pulses and signal, denoted by $\tau$ and $t$ in Fig.~\ref{fig:1}(a), enabling one to resolve excitation $\omega_\tau$ and detection $\omega_t$ frequencies, respectively. This leads to two-dimensional data sets in the $(\omega_\tau,\omega_t)$ domain for each time delay $T$ between excitation and detection processes, revealing multiple cross peaks centered at $\omega_\tau\neq\omega_t$, as well as diagonal peaks excited and detected at the same frequency $\omega_\tau\approx\omega_t$. The transient of a peak during waiting times $T$ typically exhibits damped oscillations reflecting the dynamics of quantum coherences.

Single-color 2D experiments consider three identical pulses with the same spectral features. In two-color experiments, however, the first two pulses, used to resolve excitation frequencies, can be tuned to be different from the third pulse, enabling one to consider different ranges of excitation and detection frequencies and therefore study the interaction between excitons that are widely separated in energy~\cite{MyersOE2008}. In other multi-color experiments, narrowband pulses have been considered to selectively induce specific transitions resonant with each pulse~\cite{LeeScience2007,WomickJCP2010,WrightARPC2011,RichardsJPCL2012,RichardsJPB2012,RichardsJPCL2014,RyuJPCB2014,TollerudOE2014,NovelliJPCL2015,SenlikJPCL2015}. This enables the generation of target coherences, even though ground state coherences can be induced by pulse sequences directly or mediated by finite pulse effects~\cite{refSM}. Here we consider a multi-color scheme based on broadband pulses (see Fig.~\ref{fig:1}(b)), where the blue-shift of the second pulse with respect to the first pulse suppresses the generation of the ground state coherence at all the diagonal and cross peaks within the excitation and detection windows.

To explain the principle of the proposed scheme, we consider a two-site system where site $k$ is described by its electronic ground $\ket{g_k}$ and excited $\ket{e_k}$ states
\begin{equation}
	H_s=E_1\sigma_1^\dagger \sigma_1+E_2\sigma_2^\dagger \sigma_2 + J_{12}(\sigma_1^\dagger \sigma_2 +\sigma_2^\dagger \sigma_1),
	\label{eq:dimer}
\end{equation}
where $\sigma_k^\dagger=\ket{e_k}\bra{g_k}$ denotes the raising operator of an electronic excitation at site $k$. Electronic coupling $J_{12}$ between sites makes excitons $\ket{\epsilon_k}$, namely the eigenstates of $H_s$ in the single excitation subspace, $H_s \ket{\epsilon_k}=\epsilon_k \ket{\epsilon_k}$, to be delocalized over two sites. The global ground state $\ket{g}=\ket{g_1,g_2}$ and doubly excited state $\ket{f}=\ket{e_1,e_2}$ are also the eigenstates of system Hamiltonian $H_s$ (see Fig.~\ref{fig:1}(c)). Motivated by actual PPCs and synthetic organic molecules~\cite{JordanidesJPCB2004,HayesScience2013,DiehlJPCL2014,PajusaluPRE2015}, the detuning between sites, $\Delta E=E_2-E_1$, can be present with the magnitude up to $\sim 1000\,{\rm cm}^{-1}$, which is comparable to the electronic coupling strength $J_{12}$. In this work, we consider $\Delta E=700\,{\rm cm}^{-1}$ and $J_{12}=200\,{\rm cm}^{-1}$, as model parameters, and assume that the transition dipole moments of two sites are parallel for the simplicity of 2D simulations. We employ the Franck-Condon approximation and do not involve non-adiabatic processes like internal conversion between different electronic excitation manifolds~\cite{Albert2015}.

In many PPCs, electronic couplings are comparable in magnitude to coupling to the environment, which invalidates the perturbative description of any of these couplings. In this work, we employ hierarchical equations of motion (HEOM)~\cite{KreisbeckJPCL2012,refSM,KuboJPSJ1989,TanimuraJPSJ2006}, which enables one to compute electronic-vibrational dynamics in a numerically exact manner without any perturbative treatments. We consider local phonon environments at room temperature $T=300\,K$ where phonon spectral densities are modelled by a sum of a sharp Lorentzian peak and a broad Ohmic peak, modelling underdamped intra-pigment modes and noise-inducing protein/solvent motions, respectively. In the simulations, typical values of PPCs are considered~\cite{PieperJPCB2011,IshizakiPNAS2009}. Namely, the Lorentzian peak is modelled by a Huang-Rhys factor of 0.05, quantifying vibronic coupling strength, with the mode damping time of $1\,{\rm ps}$, and vibrational frequency $\nu$ resonant with the exciton splitting, $\Delta\epsilon_{21}=\epsilon_2-\epsilon_1\approx 800\,{\rm cm}^{-1}$, which is higher than thermal energy $k_B T\approx 200\,{\rm cm}^{-1}$ at room temperature. The Ohmic part of the spectrum is modelled by the reorganization energy of $50\,{\rm cm}^{-1}$ and bath relaxation time of $100\,{\rm fs}$.

%%%%%%%%%%%%%%% Figure 1

The energy-level structure of a vibronic model, determining the frequencies of optical transitions and oscillatory 2D signals, can be well described by the eigenstates of a vibronic Hamiltonian where underdamped vibrational modes are included as a part of the system Hamiltonian. As schematically shown in Fig.~\ref{fig:1}(d), the electronic ground state manifold consists of ground states $\ket{g,n}$ with $n$ vibrational excitations, while the excited state manifold is comprised of vibronic eigenstates $\ket{\psi_k}$ which involve a coherent mixing of different excitons mediated by vibronic resonance~\cite{PriorPRL2010,ChristenssonJPCB2012,KreisbeckJPCL2012,ChinNP2013,Plenio2013,TiwariPNAS2013,ChenuSR2013,ButkusJCP2014,LimNC2015,DeSioNC2016}. For instance, the resonance between $\ket{\epsilon_1,1}$ and $\ket{\epsilon_2,0}$ and vibronic coupling between them make their superpositions to be vibronic eigenstates $\ket{\psi_{k=2,3}}$ (see Fig.~\ref{fig:1}(d)).

Here we start with a qualitative explanation of the effect of the multi-color pulses on 2D spectra before examining simulated results in detail. The generation of ground state coherences in 2D spectra is described in Fig.~\ref{fig:1}(e). In case that the thermal populations of underdamped modes in their equilibrium states are negligible due to a sufficiently high vibrational frequency, $\nu>k_B T$, the initial state is well described by the global ground state $\ket{g,0}\bra{g,0}$. Ground state coherences, for instance $\ket{g,0}\bra{g,1}$, are generated when the first pulse induces optical transition from $\bra{g,0}$ to $\bra{\psi_k}$, and then the second pulse induces the transition from $\bra{\psi_k}$ to a vibrationally excited ground state $\bra{g,1}$. This is possible when the frequencies of both optical transitions are within the laser spectrum, as shown in black in Fig.~\ref{fig:1}(e). This implies that the generation of ground state coherences can be suppressed by blue-shifting the second pulse, such that the second optical transition becomes non-resonant with the laser spectrum, as shown in blue. This is contrary to the case of a red-shift, which can still induce optical transition to vibrationally excited ground states, as shown in red.

On the other hand, when vibrational frequencies are comparable to thermal energy, the equilibrium state is a Boltzmann distribution of the global ground state $\ket{g,0}\bra{g,0}$ and vibrationally excited ground states $\ket{g,n}\bra{g,n}$ with $n\ge 1$. In this case, ground state coherences cannot be fully suppressed by the blue-shifted second pulse. For instance, starting from $\ket{g,1}\bra{g,1}$, the first two pulses can induce optical transitions from $\bra{g,1}$ to $\bra{\psi_k}$, and then to $\bra{g,0}$, as shown in Fig.~\ref{fig:1}(f). Here the second transition frequency is higher than the first one, contrary to Fig.~\ref{fig:1}(e), which is still covered by the laser spectrum of the second pulse shown in blue. Therefore the blue-shift of the second pulse is not effective to suppress such a ground state coherence $\ket{g,1}\bra{g,0}$, although the dominant component of ground state coherences stemming from $\ket{g,0}\bra{g,0}$ can be still suppressed by the scheme. This implies that our scheme cannot efficiently suppress ground state coherences for systems with small excitonic gaps~\cite{refSM}, such as the FMO complex.

The effect on the excited state signals is however very different and excited state coherences are not fully suppressed by the blue-shift of the second pulse. Starting from the global ground state, a vibronic coherence $\ket{\psi_2}\bra{\psi_1}$ can be generated via a first transition from $\bra{g,0}$ to $\bra{\psi_1}$, followed by a second transition from $\ket{g,0}$ to $\ket{\psi_2}$. In the case that $\ket{\psi_2}$ is higher in energy than $\ket{\psi_1}$, such an excited state coherence can be generated by the blue-shifted second pulse, as shown in Fig.~\ref{fig:1}(g). On the other hand, the generation of $\ket{\psi_1}\bra{\psi_2}$ can be suppressed by the blue-shift, as the second transition from $\ket{g,0}$ to $\ket{\psi_1}$ becomes non-resonant with the second pulse. This implies that the blue-shift can suppress only partially vibronic coherences, while most of the ground state coherences originating from high frequency modes are suppressed, thus ensuring that the oscillatory 2D signals are dominated by excited state coherences. This qualitative discussion to understand the rationale behind our approach is supplemented in the SM with a full analysis of the theoretical nonlinear response~\cite{refSM}.

%%%%%%%%%%%%%%% Figure 2
\begin{figure}
	\includegraphics[width=0.45\textwidth]{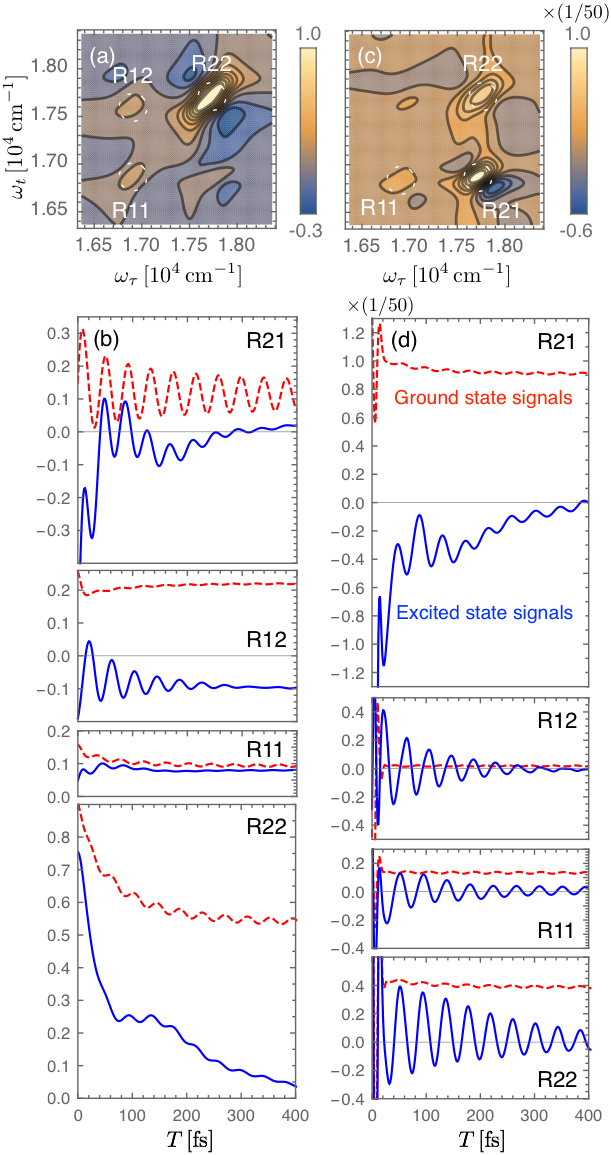}
	\caption{2D spectral response of a model vibronic dimer. (a-b) Rephasing spectra obtained by single-color scheme: (a) 2D lineshape at $T\approx 100\,{\rm fs}$ and (b) the transients of peaks R11, R12, R21, R22 marked in (a) and (c). Excited and ground state signals are shown in blue solid and red dashed lines, respectively. (c-d) Rephasing spectra obtained by the multi-color scheme. Compared to the single-color scheme, the amplitude of the ground state oscillations are significantly reduced compared to those of the excited state and the non-oscillatory component of the total signal is strongly suppressed. Hence, while there is a 50-fold reduction in overall signal intensity, the ratio of signal to noise for the oscillatory excited state component is improved compared to the single-color scheme and not contaminated by ground state oscillations.}
	\label{fig:2}
\end{figure}

To demonstrate the performance of the multi-color scheme, Fig.~\ref{fig:2} shows HEOM calculations of rephasing 2D spectra of the vibronic dimer defined in Eq.~(\ref{eq:dimer}). Fig.~\ref{fig:2}(a) displays the 2D lineshape at $T=100\,{\rm fs}$, obtained by the standard single-color scheme where Gaussian pulses with the pulse duration of $10\,{\rm fs}$ are considered, corresponding to a FWHM of $\sim 1500\,{\rm cm}^{-1}$. The central frequencies of all the pulses are taken to be $\Omega_{k=1,2,3} \approx 1.74\times 10^{4}\,{\rm cm}^{-1}$, which is the average of site energies $(E_1+E_2)/2$, so that the laser spectrum can induce optical transitions from $\ket{g,0}$ to $\ket{\psi_{k=1,2,3}}$. Due to the relatively small energy-gap between $\ket{\psi_{2}}$ and $\ket{\psi_{3}}$ compared to noise-induced homogeneous broadening, multiple peaks associated with $\ket{\psi_{k=2,3}}$ are merged to a seemingly single diagonal peak, denoted by R22. For the four peaks R11, R12, R21, R22 marked in Fig.~\ref{fig:2}(a), Fig.~\ref{fig:2}(b) displays their transients during waiting times $T$, where excited and ground state signals are shown as solid and dashed lines, respectively. Here the ground state signals include the ground state bleaching components of rephasing and non-rephasing pathways, while the excited state signals include the stimulated emission and excited state absorption components of rephasing and non-rephasing pathways, as well as double-quantum pathways, contributing to rephasing 2D spectra mediated by finite pulse effects~\cite{BrixnerJCP2004,refSM}. The damping of the oscillation amplitude of ground state signals is hardly visible, as the damping time of underdamped modes is taken to be $1\,{\rm ps}$, while the oscillatory features in excited state signals disappear more rapidly due to the dephasing induced by excitonic characters. It is worth noting that in 2D experiments, only the sum of the ground and excited state signals can be measured, implying that coherent features in the excited state cannot be directly measured when 2D spectra are contaminated by ground-state coherences.

In the multi-color case, the central frequency of the first pulse is fixed to be $\Omega_1\approx 1.74\times 10^{4}\,{\rm cm}^{-1}$, while those of the second and third pulses are taken to be $\Omega_2=\Omega_1+\Delta\Omega$ and $\Omega_3=\Omega_1-\Delta\Omega$, respectively, where the third pulse is red-shifted in order to make the detection window coincide with the excitation window~\cite{refSM}. As $\Delta\Omega$ increases, the oscillation amplitude of the ground state signals is reduced relative to the excited state signals at all the peak locations, and that the oscillatory features in rephasing spectra start to be dominated by excited state coherences as $\Delta\Omega$ becomes of the order of vibrational frequency $\nu$. As $\Delta\Omega$ increases further, the overall intensity of 2D spectra is reduced with the ground state coherences further suppressed. While the overall signal intensity is reduced with increasing $\Delta\Omega$, we would like to stress that the signal to noise ratio for detection of the excited state oscillatory signal is improved relative to the single-color scheme as both the ground state oscillations are suppressed and the non-oscillatory background is strongly reduced compared to the single-color scheme. In Fig.~\ref{fig:2}(c), 2D lineshape at $T=100\,{\rm fs}$ is displayed for $\Delta\Omega=2000\,{\rm cm}^{-1}$, where a below-diagonal cross peak R21 becomes more visible than the single-color case. The transients of the four peaks in Fig.~\ref{fig:2}(d) demonstrate that the oscillatory features in ground state signals can be efficiently suppressed by the multi-color scheme, enabling the direct observation of excited state coherences from raw 2D spectra. Short-lived oscillations of ground state signals up to $T\approx 30\,{\rm fs}$ are induced by the overlap between pulses with different colors, and such features also appear in the excited state signals. We note that the vibronic coupling to underdamped modes is essential to observe long-lived 2D oscillations in Fig.~\ref{fig:2}, implying that the coherent features are induced by intrinsic vibronic dynamics, rather than being an artefact caused by the coherent laser fields themselves~\cite{refSM}.

So far we have demonstrated that the multi-color scheme can suppress ground state coherences for a dimer system when underdamped modes are resonant with exciton splitting. In PPCs, however, electronic parameters are not well defined due to static disorder, and multiple underdamped modes can be present with different vibrational frequencies. In the SM, we show that our scheme can efficiently suppress the ground state coherences for dimer models including static disorder and multiple modes with different vibrational frequencies that are quasi-resonant with exciton splitting. In addition, we demonstrate the performance of our scheme for a photosynthetic complex model consisting of eight pigments, based on the parameters of the phycocyanin 645 from marine algae \cite{ColliniN2010,Aspuru2018}, to highlight the feasibility of our scheme for realistic multi-chromophore systems.

In summary, we proposed a multi-color scheme for suppressing ground state coherences in 2D electronic spectroscopy. We stress that our scheme enables raw 2D data to be dominated by coherent excited state dynamics, but a detailed analysis is still required to interpret the origin of the isolated 2D oscillations, as is the case of the standard 2D experiments. We note that broadband pulses are essential to suppress ground state coherences, as spectrally-narrow long pulses can lead to finite pulse-duration effects, which can induce ground state coherences even for a blue-shifted second pulse~\cite{refSM}. It is notable that with the development of pulse shapers, it is possible to modulate various properties of laser pulses, such as amplitude, phase and polarization, for a sub-$10\,{\rm fs}$ light source~\cite{MaOE2016,DraegerOE2017,MuellerJPCL2018}. This suggests the possibility to further improve our scheme by controlling other properties of pulses, such as chirping. Such developments may be helpful for the actual experimental implementation and eventually for the unambiguous identification of coherent excited state dynamics in PPCs and organic photovoltaics~\cite{DeSioNC2016,FalkeS2014}.

We thank Jinhyoung Lee, Tobias Kramer and J{\" u}rgen Hauer for insightful discussions. This work was supported by  the ERC Synergy grant BioQ and the John Templeton Foundation.

%%%%%%%%%%%%%%%
%%%%%%%%%%%%%%% References
%%%%%%%%%%%%%%%


\begin{thebibliography}{99}

\bibitem{MayKuehn2003} V. May and O. K\"{u}hn, Charge and Energy Transfer Dynamics in Molecular Systems (Wiley-VCH, 2003).

\bibitem{forster1946} T. F\"{o}rster,  Naturwissenschaften \textbf{33}, 166 (1946).

\bibitem{Amerongen2000} H. van Amerongen, L. Valkunas, and R. van Grondelle, Photosynthetic Excitons (World Scientific, 2000).

\bibitem{Valkunas} L. Valkunas, D. Abramavicius, and T. Man{\v c}al, Molecular Excitation Dynamics and Relaxation: Quantum Theory and Spectroscopy (Wiley-VCH, 2013).

\bibitem{JonasARPC2003} D. M. Jonas, Annu. Rev. Phys. Chem. \textbf{54}, 425 (2003).

\bibitem{BrixnerJCP2004} T. Brixner, T. Man{\v c}al, I. V. Stiopkin, and G. R. Fleming, J. Chem. Phys. \textbf{121}, 4221 (2004).

\bibitem{Blankenship2002} R. E. Blankenship, Molecular Mechanism of Photosynthesis (World Scientific, 2002).

\bibitem{EngelN2007} G. S. Engel, T. R. Calhoun, E. L. Read, T.-K. Ahn, T. Man{\v c}al, Y.-C. Cheng, R. E. Blankenship, and G. R. Fleming, Nature \textbf{446}, 782 (2007).

\bibitem{EngelPNAS2010} G. Panitchayangkoon, D. Hayes, K. A. Fransted, J. R. Caram, E. Harel, J. Wen, R. E. Blankenship, and G. S. Engel, Proc. Natl. Acad. Sci. U. S. A. \textbf{107}, 12766 (2010).

\bibitem{ColliniN2010} E. Collini, C. Y. Wong, K. E. Wilk, P. M. G. Curmi, P. Brumer, and G. D. Scholes, Nature \textbf{463}, 644 (2010).

\bibitem{RomeroNP2014} E. Romero, R. Augulis, V. I. Novoderezhkin, M. Ferretti, J. Thieme, D. Zigmantas, and R. van Grondelle, Nat. Phys. \textbf{10}, 676 (2014).

\bibitem{FullerNP2014} F. D. Fuller, J. Pan, A. Gelzinis, V. Butkus, S. S. Senlik, D. E. Wilcox, C. F. Yocum, L. Valkunas, D. Abramavicius, and J. P. Ogilvie, Nat. Chem. \textbf{6}, 706 (2014).

\bibitem{Huelga2013} S. F. Huelga and M. B. Plenio, Contemp. Phys. \textbf{54}, 181 (2013).

\bibitem{Jonas2018}  D. M. Jonas, Annu. Rev. Phys. Chem. \textbf{69}, 327 (2018).

\bibitem{rmp2018} S. J. Jang and B. Mennucci, Rev. Mod. Phys. \textbf{90}, 035003 (2018).

\bibitem{PriorPRL2010} J. Prior, A. W. Chin, S. F. Huelga, and M. B. Plenio, Phys. Rev. Lett. \textbf{105}, 050404 (2010).

\bibitem{ChristenssonJPCB2012} N. Christensson, H. F. Kauffmann, T. Pullerits, and T. Man{\v c}al, J. Phys. Chem. B \textbf{116}, 7449 (2012).

\bibitem{KreisbeckJPCL2012} C. Kreisbeck and T. Kramer, J. Phys. Chem. Lett. \textbf{3}, 2828 (2012).

\bibitem{ChinNP2013} A. W. Chin, J. Prior, R. Rosenbach, F. Caycedo-Soler, S. F. Huelga, and M. B. Plenio, Nat. Phys. \textbf{9}, 113 (2013).

\bibitem{Plenio2013} M. B. Plenio, J. Almeida, and S. F. Huelga, J. Chem. Phys. \textbf{139}, 235102 (2013).

\bibitem{TiwariPNAS2013} V. Tiwari, W. K. Peters, and D. M. Jonas, Proc. Natl. Acad. Sci. U. S. A. \textbf{110}, 1203 (2013).

\bibitem{ChenuSR2013} A. Chenu, N. Christensson, H. F. Kauffmann, and T. Man{\v c}al, Sci. Rep. \textbf{3}, 2029 (2013).

\bibitem{ButkusJCP2014} V. Butkus, L. Valkunas, and D. Abramavicius, J. Chem. Phys. \textbf{140}, 034306 (2014).

\bibitem{LimNC2015} J. Lim, D. Pale{\v c}ek, F. Caycedo-Soler, C. N. Lincoln, J. Prior, H. von Berlepsch, S. F. Huelga, M. B. Plenio, D. Zigmantas, and J. Hauer, Nat. Commun. \textbf{6}, 7755 (2015).

\bibitem{DeSioNC2016} A. De Sio, F. Troiani, M. Maiuri, J. R{\' e}hault, E. Sommer, J. Lim, S. F. Huelga, M. B. Plenio, C. A. Rozzi, G. Cerullo, E. Molinari, and C. Lienau, Nat. Commun. \textbf{7}, 13742 (2016).

\bibitem{Donatas2018} E. Thyrhaug, R. Tempelaar, M. J. P. Alcocer, K. {\v Z}{\' i}dek, D. B{\' i}na, J. Knoester, T. L. C. Jansen, and D. Zigmantas, Nat. Chem. \textbf{10}, 780 (2018).

\bibitem{Maiuri2018} M. Maiuri, E. E. Ostroumov, R. G. Saer, R. E. Blankenship and G. D. Scholes, Nat. Chem. \textbf{10}, 177 (2018).

\bibitem{EngelChem2018} B. S. Rolczynski, H. Zheng, V. P. Singh, P. Navotnaya, A. R. Ginzburg, J. R. Caram, K. Ashraf, A. T. Gardiner, S.-H. Yeh, S. Kais, R. J. Cogdell, and G. S. Engel, Chem \textbf{4}, 138 (2018).

\bibitem{Miller2017} H.-G. Duan, V. I. Prokhorenko, R. J. Cogdell, K. Ashraf, A. L. Stevens, M. Thorwart, and R. J. Dwayne Miller, Proc. Natl. Acad. Sci. U. S. A. \textbf{114}, 8493 (2017).

\bibitem{Scholes2016} J. C. Dean, T. Mirkovic, Z. S. D. Zoa, D. G. Oblinsky and G. D. Scholes, Chem. \textbf{1}, 858 (2016).

\bibitem{Scholes2018} C. C. Jumper, I. H. M. van Stokkum, T. Mirkovic and G. D. Scholes, J. Phys. Chem. B \textbf{122}, 6328 (2018).

\bibitem{Kolli2012} A. Kolli, E. J. O'Reilly, G. D. Scholes, and  A. Olaya-Castro, J. Chem. Phys. \textbf{137}, 17, 174109 (2012).

\bibitem{Aspuru2018} S. M. Blau, D. I. G. Bennett, C. Kreisbeck, G. D. Scholes, and A. Aspuru-Guzik, Proc. Natl. Acad. Sci. U. S. A. \textbf{115}, E3342 (2018).

% P-controlled

\bibitem{Donatas2012} S. Westenhoff, D. Pale{\v c}ek, P. Edlund, P. Smith, and D. Zigmantas, J. Am. Chem. Soc. \textbf{134}, 16484 (2012).

% SM

\bibitem{refSM} See Supplemental Material for the comparison of quantum control schemes proposed to suppress ground state coherences in 2D electronic spectroscopy, and theoretical methods used in this work, which includes Refs.~\cite{JonasARPC2003,Valkunas,BrixnerJCP2004,SMMukamel,PriorPRL2010,TiwariPNAS2013,Plenio2013,ChinNP2013,ChristenssonJPCB2012,LimNC2015,KuboJPSJ1989,TanimuraJPSJ2006,KreisbeckJPCL2012,SenlikJPCL2015,Donatas2012,Donatas2018,ColliniN2010,SMMirkovic2007,Aspuru2018,SMSpano,DeSioNC2016,WSCP,Joel,SMJoel2}.

\bibitem{SMMukamel} S. Mukamel, Principles of Nonlinear Optical Spectroscopy (Oxford University Press, 1995).

\bibitem{SMMirkovic2007} T. Mirkovic, A. B. Doust, J. Kim, K. E. Wilk, C. Curutchet, B. Mennucci, R. Cammi, P. M. G. Curmib, and G. D. Scholes, Photochem. Photobiol. Sci. {\textbf 6}, 964 (2007).

\bibitem{SMSpano} F. C. Spano, J. Chem. Phys. {\bf 116}, 5877 (2002).

\bibitem{WSCP} T.-C. Dinh and T. Renger, J. Chem. Phys. \textbf{142}, 034104 (2015).

\bibitem{SMJoel2} J. Yuen-Zhou, A. S. Johnson, J. J. Krich, A. Aspuru-Guzik, and Ivan Kassal, Ultrafast Spectroscopy: Quantum information and wavepackets (IOP Publishing, Bristol, 2014).

% Broadband PP/TG

\bibitem{Joel} J. Yuen-Zhou, J. J. Krich, and A. Aspuru-Guzik, J. Chem. Phys. {\bf 136}, 234501 (2012).

\bibitem{Joel2} A. S. Johnson, J. Yuen-Zhou, A. Aspuru-Guzik, and J. J. Krich, J. Chem. Phys. {\bf 141}, 244109 (2014).

% Omega1 = Omega2

\bibitem{MyersOE2008} J. A. Myers, K. L. M. Lewis, P. F. Tekavec, and J. P. Ogilvie, Opt. Express \textbf{16}, 17420 (2008).

% Omega1 != Omega2

\bibitem{LeeScience2007} H. Lee, Y. C. Cheng, and G. R. Fleming, Science \textbf{316}, 1462 (2007).

\bibitem{WomickJCP2010} J. M. Womick, S. A. Miller, and A. M. Moran, J. Chem. Phys. \textbf{133}, 024507 (2010).

\bibitem{WrightARPC2011} J. C. Wright, Annu. Rev. Phys. Chem. \textbf{62}, 209 (2011).

\bibitem{RichardsJPCL2012} G. H. Richards, K. E. Wilk, P. M. G. Curmi, H. M. Quiney, and J. A. Davis, J. Phys. Chem. Lett. \textbf{3}, 272 (2012).

\bibitem{RichardsJPB2012} G. H. Richards, K. E. Wilk, P. M. G. Curmi, H. M. Quiney, and J. A. Davis, J. Phys. B: At. Mol. Opt. Phys. \textbf{45}, 154015 (2012).

\bibitem{RichardsJPCL2014} G. H. Richards, K. E. Wilk, P. M. G. Curmi, and J. A. Davis, J. Phys. Chem. Lett. \textbf{5}, 43 (2014).

\bibitem{RyuJPCB2014} I. S. Ryu, H. Dong, and G. R. Fleming, J. Phys. Chem. B \textbf{118}, 1381 (2014).

\bibitem{TollerudOE2014} J. O. Tollerud, C. R. Hall, and J. A. Davis, Opt. Express \textbf{22}, 6719 (2014).

\bibitem{NovelliJPCL2015} F. Novelli, A. Nazir, G. H. Richards, A. Roozbeh, K. E. Wilk, P. M. G. Curmi, and J. A. Davis, J. Phys. Chem. Lett. \textbf{6}, 4573 (2015).

\bibitem{SenlikJPCL2015} S. S. Senlik, V. R. Policht, and J. P. Ogilvie, J. Phys. Chem. Lett. \textbf{6}, 2413 (2015).

% Dimer parameters

\bibitem{JordanidesJPCB2004} X. J. Jordanides, G. D. Scholes, W. A. Shapley, J. R. Reimers, and G. R. Fleming, J. Phys. Chem. B \textbf{108}, 1753 (2004).

\bibitem{HayesScience2013} D. Hayes, G. B. Griffin, G. S. Engel, Science {\bf 340}, 1431 (2013).

\bibitem{DiehlJPCL2014} F. P. Diehl, C. Roos, A. Duymaz, B. Lunkenheimer, A. K{\"o}hn, and T. Basche, J. Phys. Chem. Lett. \textbf{5} 262 (2014).

\bibitem{PajusaluPRE2015} M. Pajusalu, R. Kunz, M. R{\"a}tsep, K. Timpmann, J. K{\"o}hler, and A. Freiberg, Phys. Rev. E \textbf{92}, 052709 (2015).

% Internal conversion

\bibitem{Albert2015} J. Albert, M. Falge, S. Gomez, I. R. Sola, H. Hildenbrand, and V. Engel, J. Chem. Phys. \textbf{143}, 041102 (2015).

% HEOM

\bibitem{KuboJPSJ1989} Y. Tanimura and R. Kubo, J. Phys. Soc. Jpn. \textbf{58}, 101 (1989).

\bibitem{TanimuraJPSJ2006} Y. Tanimura, J. Phys. Soc. Jpn. \textbf{75}, 082001 (2006).

% HR and reorganization energy

\bibitem{PieperJPCB2011} J. Pieper, M. R{\"a}tsep, I. Trostmann, H. Paulsen, G. Renger, and A. Freiberg, J. Phys. Chem. B \textbf{115}, 4042 (2011).

\bibitem{IshizakiPNAS2009} A. Ishizaki and G. R. Fleming, Proc. Natl. Acad. Sci. U. S. A. \textbf{106}, 17255 (2009).

% Pulse shapers

\bibitem{MaOE2016} X. Ma, J. Dost{\'a}l, and T. Brixner, Opt. Express \textbf{24}, 20781 (2016).

\bibitem{DraegerOE2017} S. Draeger, S. Roeding, and T. Brixner, Opt. Express \textbf{25}, 3259 (2017).

\bibitem{MuellerJPCL2018} S. Mueller, S. Draeger, X. Ma, M. Hensen, T. Kenneweg, W. Pfeiffer, and T. Brixner, J. Phys. Chem. Lett. \textbf{9}, 1964 (2018).

% OPV

\bibitem{FalkeS2014} S. M. Falke, C. A. Rozzi, D. Brida, M. Maiuri, M. Amato, E. Sommer, A. De Sio, A. Rubio, G. Cerullo, E. Molinari, and C. Lienau, Science \textbf{344}, 1001 (2014).

\end{thebibliography}
\end{document}